\documentclass[10pt,twocolumn,prb,reqno, amsmath,amssymb, aps]{revtex4-1}

\usepackage[pdftex]{graphicx}
\usepackage{float}

\usepackage{perpage}

\usepackage{color}
\usepackage{rotating}
\usepackage[breaklinks=true,colorlinks,citecolor=blue,linkcolor=blue,urlcolor=blue]{hyperref}
\usepackage{appendix}
\usepackage{graphicx}
\usepackage{natbib}
\usepackage{dcolumn}
\usepackage{bm}
\usepackage{hyperref}
\usepackage[mathlines]{lineno}
\usepackage[caption=false]{subfig}
\usepackage{epstopdf}

\begin{document}

\title{Strong and tunable couplings in flux-mediated optomechanics}
\author{Olga Shevchuk, Gary A. Steele, and  Ya. M. Blanter}
\affiliation{Kavli Institute of Nanoscience, Delft University of Technology, Lorentzweg 1, 2628 CJ Delft, The Netherlands}

\begin{abstract}
We investigate superconducting interference device (SQUID) with two asymmetric Josephson junctions coupled to a mechanical resonator embedded in the loop of the SQUID. We quantize this system in the case when the frequency of the mechanical resonator is much lower than the cavity frequency of the SQUID and in the case when they are comparable. In the first case, the radiation pressure and cross-Kerr type interactions arise and are modified by asymmetry. Cross-Kerr type coupling is the leading term at the extremum points where radiation pressure is zero.  In the second case, the main interaction is single-photon beam splitter, which exists only at finite asymmetry.  Another interaction in this regime is of cross-Kerr type, which exists at all asymmetries, but generally much weaker than the beam splitter interaction. Increasing magnetic field can substantially enhance optomechanical couplings strength with the potential for the radiation pressure coupling to reach the single-photon strong coupling regime, even the ultrastrong coupling regime, in which the single-photon coupling rate exceeds the mechanical frequency.
\end{abstract}
\maketitle
\section{Introduction}

The progress in optomechanical systems, where an optical or microwave cavity is coupled to a mechanical resonator, was impressive in recent years\cite{Aspelmeyer}. The accomplishments in optomechanics include cooling mechanical resonator to its quantum ground state\cite{Chan, Teufel}, prediction\cite{OMIT1} and observation\cite{OMIT2} of the optomechanically-induced transparency, squeezing of the cavity\cite{squeezing1,squeezing2}, and mechanical\cite{squeezing3,squeezing4,squeezing5} modes, and coherent state transfer\cite{statestransfer1, statestransfer2}. Many of these experiments have been realized using superconducting circuits, which enables to consider microwave cavities coupled to mechanical motion as possible building blocks for quantum information processing\cite{QIP}.

The coupling between cavity and mechanical resonator plays a central role in optomechanics. In the published experiments, intrinsically weak radiation pressure coupling was amplified by increasing drive power of the cavity, which linearizes the effective optomechanical interaction of the system. Such linear interaction, for example, turns Gaussian states of the cavity and mechanical resonator into Gaussian states. In order to create more general states for quantum information applications and to achieve, for instance, negative Wigner function one needs to use either single-photon sources and photodetectors\cite{Knill} or non-linear effects, of which non-linear optomechanical interaction is the most common one. Therefore, having strong single-photon radiation pressure coupling of the order or larger than the cavity decay rate is desirable as well as having strong coupling of the cavity to the position squared of the mechanical resonator\cite{Harris_one}. If the single-photon radiation pressure coupling can be made of the order of the mechanical frequency and larger than the cavity decay rate, the the system is in the ultra-strong coupling regime and photon blockade can be observed\cite{Rabl}. 

Along with the ultracold atoms\cite{Stamper-Kurn}, superconducting circuits are promising candidates to reach ultrastrong coupling. Recently, the idea of using Josephson effect to enhance optomechanical couplings has been researched theoretically\cite{Nori, Sillanpaa, Blencowe} and experimentally\cite{Sillanpaa_exp}. Many of those proposals involve using superconducting quantum interference device (SQUID) with two Josephson junctions, which makes cavity intrinsically nonlinear due to the Josephson effect. SQUID is either embedded into the resonator itself or SQUID with embedded mechanical resonator is incorporated into a microwave cavity. 

In this Article, we consider a SQUID with  two symmetric or asymmetric Josephson junctions and an embedded mechanical resonator and show that it by itself can produce ultrastrong optomechanical coupling.  Originally, a dc SQUID with embedded mechanical oscillator was studied as a sensitive displacement detector
\cite{Zhou, Blencowe_2006, Pugnetti,Etaki,Etaki1}, however, the asymmetry of the junsctions so far was not at the focus of attention, and theoretical proposals are routinely assuming that two junctions of the SQUID are almost identical.  A certain asymmetry is always present in the experiment, and we show that it affects the coupling strength.  In addition, we express the couplings in such SQUID devices in the language of optomechanics, perform numerical simulations of the coupling rates for realistic experimental geometries. Doing so we find that this platform has the potential to reach both the single-photon strong coupling, a regime of strong quadratic coupling of the motion to the cavity, and potentially the ultrastrong coupling regime where the single-photon coupling rate exceeds the mechanical frequency.

In the first part of the article, we investigate in details the effect of asymmetry in the SQUID with two junctions and embedded mechanical resonator. As a first step we look at the most common experimental case of the mechanical frequency being much smaller than the cavity frequency\cite{vanderZant}. We quantize the asymmetric system to get radiation pressure interaction and cross-Kerr type interaction, where the cavity is coupled to the position squared of the mechanical resonator. We show that for experimentally feasible parameters radiation pressure coupling can reach single-photon strong coupling regime and for stronger magnetic fields the ultrastrong coupling regime. The cross-Kerr coupling is usually smaller than radiation pressure coupling but it is the leading coupling at the extremum points of the flux where the radiation pressure is zero. Such strong coupling would enable a quantum non-demolition measurement of a phonon number in the mechanical resonator\cite{Cross} or the cavity's photon number.  

As a second step, we study the case when the mechanical and cavity frequencies are of the same order. Since the SQUID cavity frequency is measuring in GHz, the same order would be required for the mechanical oscillator. Currently, carbon nanotube (CNT) resonator can reach GHz frequency\cite{CNT}  and, hence, the realizations of the SQUID with suspended CNT junctions\cite{CNT-SQUID, Ben} could reach this regime.  In this case, there are two leading interactions: cross-Kerr and single-photon beam splitter.  The single-photon beam splitter exists only at the finite asymmetry. The radiation pressure term is oscillating too fast and is, therefore, disregarded. The beam splitter is used in many exrimental setups, and Hamiltonian with the beam splitter interaction is easily diagonalized and solved. When single-photon beam splitter is in the range of the strong coupling, one can observe {\em e.g.} optomechanical normal-mode splitting\cite{Aspelmeyer}. 

The remainder of the Article is organized as follows. In Sec. \ref{sec:current} we find current and cavity frequency of the SQUID with asymmetric Josephson junctions and an embedded mechanical resonator. In Sec. \ref{sec:quantization} we derive the effective Hamiltonian of this system for two cases. In the first case, the cavity frequency of the SQUID is taken to be much larger than the mechanical frequency, which  results in the radiation pressure and cross-Kerr interactions. In the second case, the cavity frequency is considered to be of the order of the mechanical frequency providing single-photon beam splitter and cross-Kerr interactions. In Sec. \ref{sec:dis} we draw the potential map and discuss optomechanical couplings. Finally, we conclude our results in Sec. \ref{sec:concl}.

\section{Current of the asymmetric SQUID}
\label{sec:current}
\begin{figure}
\centering
    \subfloat[]{\includegraphics[width=0.55\linewidth]{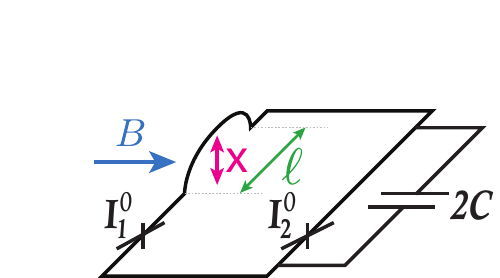}}
    \quad
        \subfloat[]{\includegraphics[width=0.4\linewidth]{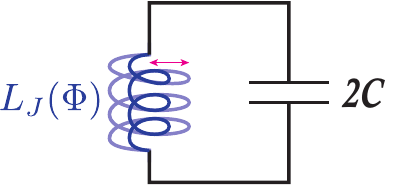}}
    \\
    \subfloat[]{\includegraphics[trim = 0 0 90 0, clip, width=\linewidth]{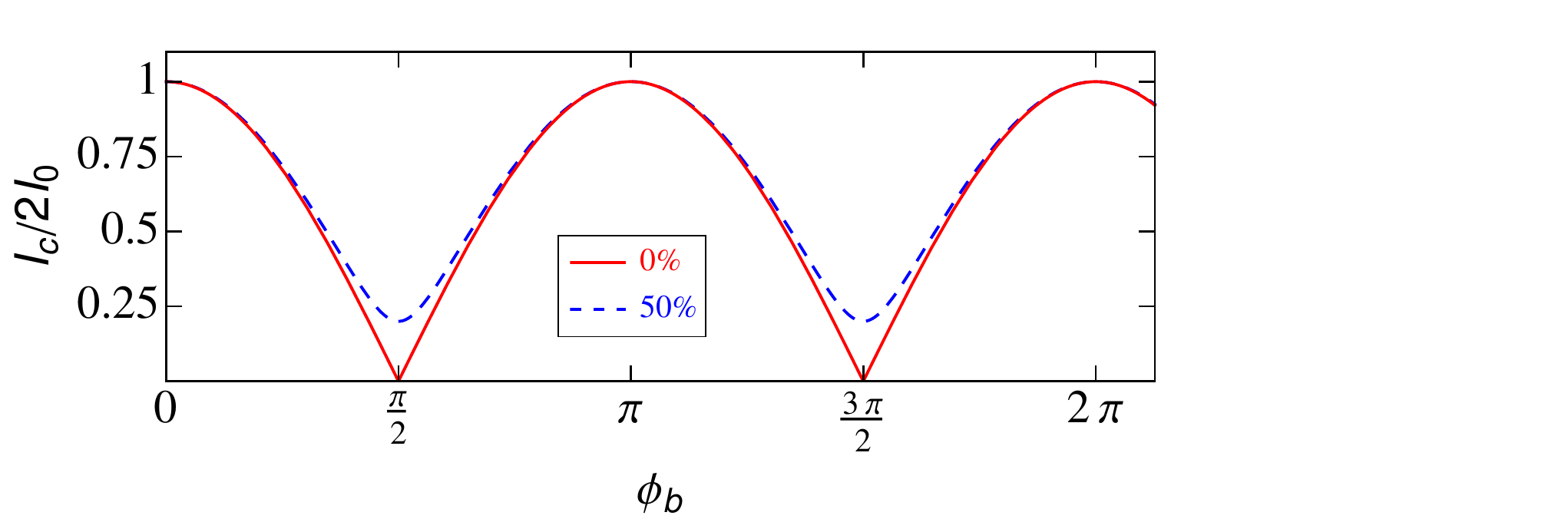}}
    \\
 \subfloat[]{\includegraphics[trim = 0 0 90 0, clip, width=\linewidth]{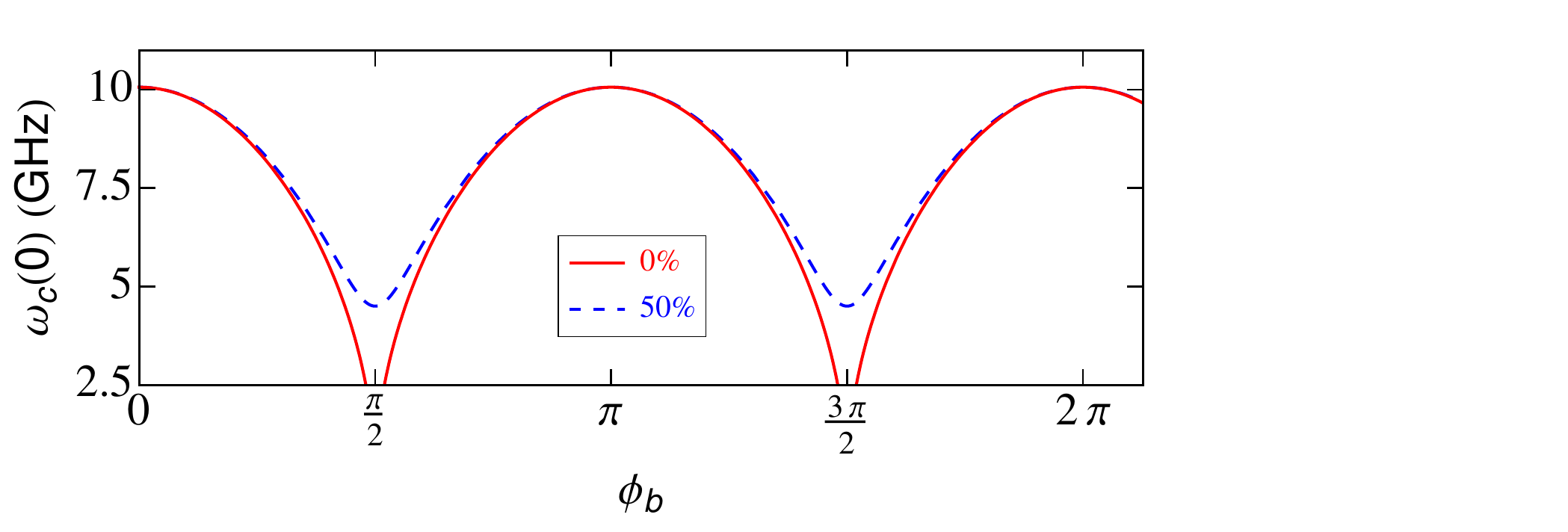}}
  \caption{(a) A schematic overview of the SQUID, which contains two Josephson junctions with different critical currents. The mechanical resonator is embedded into the SQUID loop. The magnetic field $B$ is applied under certain angle to the loop, and the displacement of the mechanical resonator is $X$. (b) The mechanical resonator inductively couples to the SQUID via the total flux $\Phi$. (c) The critical current and (d) cavity frequency are plotted as a function of renormalized bias flux of the symmetric and asymmetric SQUID. The cavity frequency of the symmetric SQUID is cut at realistic value of 2.5 GHz. }
\label{fig:freq_I}
\end{figure}

In this Section, we follow the standard textbook treatment of the current through an asymmetric SQUID. We consider two Josephson junctions with different values of critical current $I_1^0$ and $I_2^0$ connected in a loop together with an embedded mechanical resonator, as shown in Fig.~\ref{fig:freq_I}(a). The energy scales for such SQUID are described  by average Josephson energy $E_J=\hbar (I_1^0+I_2^0)/4 e$ and charging energy $E_c=(2e)^2/2C\ll E_J$ with the shunting capacitance of each junction $C$.  The SQUID has a loop area $A$ with the suspended arm of the length $l$. Oscillations of the mechanical resonator modulate the total flux of the SQUID loop. Then, the SQUID with the embedded mechanical resonator can be viewed as an LC circuit, in which the Josephson inductance of the SQUID $L_J$, which for symmetric junctions $I_1^0=I_2^0=I_0$ is well-known to be $\Phi_0/(4\pi I_0 \cos(\pi\Phi/\Phi_0))$, changes with the total flux $\Phi$ threading though the loop and, consequently, the mechanical resonator couples inductively to the SQUID, see Fig.~\ref{fig:freq_I}(b). For simplicity, we assume that mechanical resonator moves in its single mode. The dynamics of the mechanical resonator is described by the displacement $X$ from the equilibrium position. The dynamics of the SQUID itself is described by the sum of the gauge-invariant phases across each junction ($\phi_1$ and $\phi_2$), $\varphi_+=(\phi_1+\phi_2)/2$, which is referred as the overall phase of the SQUID. Moreover, the difference of the phases is bound by the total flux  threading the loop,
\begin{equation}
\varphi_-=(\phi_1-\phi_2)/2=\pi \Phi/\Phi_0+\pi n \ ,
\end{equation}
where $n$ is an integer and $\Phi_0=h/2e$ is the flux quantum.  

Assuming the magnetic field is applied under  certain angle to the SQUID loop the total flux can be separated to two contributions. The first contribution is a bias flux $\Phi_b$, which is added to the SQUID loop.  The second contribution comes from a flux threading through the area described by the oscillations of mechanical resonator. Since it is more convenient to work with the phase difference rather than the flux itself we define renormalized bias flux  $\phi_b=\pi \Phi_b/\Phi_0$ and renormalized flux shift provided by the resonator  $\xi X=\pi\beta_0 BlX/\Phi_0$ with the average geometric constant  $\beta_0$, which takes into account the direction of the magnetic field and the geometry of the mechanical resonator. Then, the phase difference is given by
\begin{equation}
\varphi_-=\phi_b+\xi X+\pi n.
\end{equation}
Here, we study the situation when the circuit has a negligible self-inductance.

Now we write the total current going through the asymmetric SQUID. For this purpose we introdice the average critical current $I_0=(I_1^0+I_2^0)/2$. The critical currents of the first and the second junctions are defined as $I_1^0=I_0(1-\alpha_I)$ and $I_1^0=I_0(1+\alpha_I)$, respectively, with the asymmetry parameter $\alpha_I$. Therefore, the total current $I$ through both junctions is separated to two terms: one is the same as in the case of equal critical currents and another term, which is responsible for the influence of asymmetry,
\begin{eqnarray}
I&&=I_1^0\sin(\phi_1)+I_2^0\sin(\phi_2)
\nonumber
\\
&&=2 I_0\cos(\varphi_-)\sin(\varphi_+)-2 I_0 \alpha_I \cos(\varphi_+)\sin(\varphi_-).
\end{eqnarray}
In order to find a critical current of the asymmetric SQUID, we  shift the position of the overall phase of the SQUID by the phase $\varphi_0$ which satisfies the relation: $\tan(\varphi_0)=\alpha_I \tan(\varphi_-)$. Then, the total current is simplified to the following
\begin{equation}
I=2I_0 S(\varphi_-)\sin(\varphi_+-\varphi_0),
\end{equation}
where $S(\varphi_-)= \sqrt{\cos^2(\varphi_-)+\alpha_I^2\sin^2(\varphi_-)}$ is a flux dependent function, which turns to cosine at zero asymmetry, and the total current becomes the well-known current of the symmetric SQUID. Then, when mechanical resonator is at rest, we can define the maximum current and, hence, the critical current of the asymmetric SQUID as well as the cavity frequency
\begin{equation}
I(X=0)=I_c=2I_0 S(0)\hspace{10pt}\text{and}\hspace{10pt}\omega_c(0)=\sqrt{\frac{2 \pi I_c}{C \Phi_0}}.
\label{omega0}
\end{equation}
Here we use $S(0)$, which is the function of $\phi_b$  instead of $\varphi_-$ at zero displacement.

In Fig.~\ref{fig:freq_I}(c) we show the behavior of the critical current for symmetric and 50\% asymmetry cases. For the identical junctions the current changes from 0 to $2I_0$, but in the presence of the asymmetry the current never reaches zero value. Even at half flux quantum when the critical current for the symmetric case is zero, the  critical current of the asymmetric SQUID is at minimum $I_c(\phi_b=\pi/2)=2I_0\alpha_I$. Nevertheless, the maximum, which happens at the odd integer flux quantum, is not affected by the asymmetry. The cavity frequency is proportional to $\sqrt{I_c}$ and portrays the same behavior of the critical current as shown in Fig.~\ref{fig:freq_I}(d). For parameters of the critical current of the Josephson junction $I_0=500$ nA and capacitance $C=30$ pF the maximum cavity frequency is 10 GHz. At half flux quantum and 50\% asymmetry, the cavity frequency reaches its minimum of 4.5 GHz.

\section{Quantization}
\label{sec:quantization}
 In the following, we quantize the system by starting with the classical Hamiltonian, which consists of the simple harmonic oscillator, kinetic energy and the potential energy of the SQUID,
\begin{align}
&H=\frac{m_r\dot{X}^2}{2}+\frac{m_r\omega_m^2 X^2}{2}+\frac{C\Phi_0^2}{2 (2\pi)^2}\dot{\varphi}^2_+ +E(\varphi_+,X),
 \label{Heff}
\end{align}
where $m_r$ and $\omega_m$ are the mass and frequency of the mechanical resonator. The potential energy of the SQUID $E$ is derived from the total current $\Phi_0 I/2\pi=\partial E/\partial \varphi_+$ found in Sec. \ref{sec:current}, 
\begin{eqnarray}
&&E(\varphi_+,X)=
-2E_J S(\varphi_-)\cos\left[\varphi_+-\arctan (\alpha_I|\tan \varphi_-|)\right].\quad
\label{Energy}
\end{eqnarray}

The minimum of the potential is shifted by the flux dependent parameter, which also depends on the displacement of the mechanical resonator. Depending on the difference between the cavity frequency and the mechanical frequency one can assume quasi-static regime or has to take into account the displacement dependent shift.

\subsection{Dispersive regime}

In the typical case when the mechanical frequency is much smaller than cavity frequency, the shift by the flux can be assumed static on the timescales related to the SQUID. Then, we can write potential energy in terms of the shifted phase, $\varphi=\varphi_+-\arctan(\alpha_I |\tan(\phi_b)|)$.   The kinetic energy of the SQUID is not affected by the constant shift, and thus the phase $\varphi_+$ can be replaced by $\varphi$. 

In order to quantize the phase and the position, the potential energy is expanded in terms of the phase  up to the second order. This means that we consider SQUID as a linear harmonic oscillator with the single-photon Kerr shift smaller than the linewidth of the cavity and the cavity frequency. The term, which is independent of the phase, shifts the equilibrium position of the mechanical resonator and modifies the mechanical frequency
\begin{equation} 
\omega_m'=\sqrt{\omega_m^2+\frac{4E_J \xi^2(1-\alpha_I^2)(\cos^4(\phi_b)-\alpha_I^2\sin^4(\phi_b))}{m_r S(0)^3}}.
\label{omega_m}
\end{equation}
For the phase dependent terms we introduce creation and annihilation operators
\begin{eqnarray}
&& \left( \begin{array}{c} a^{\dagger} \\ a \end{array} \right) =\frac{1}{\sqrt{2 \hbar m_{\varphi} \omega_c}}\left(m_{\varphi}  \omega_c\varphi\mp i p_{\varphi}\right).
\end{eqnarray}
with the momentum coordinate  $p_{\varphi}=C\Phi_0^2/ (2\pi)^2\dot{\varphi} \equiv m_{\varphi}\dot{\varphi}$, where $m_{\varphi}$ is the mass of the phase,   and the displacement dependent cavity frequency is
\begin{equation} \label{cavity_adiabatic}
\omega_c(\varphi_-)= \sqrt{\frac{4\pi I_0 S(\varphi_-) }{C\Phi_0}} \ .
\end{equation}
This expression can also be retrieved from Eq. \eqref{omega0} for the mechanical resonator at rest by changing $\phi_b$ to $\varphi_-$. Therefore, the displacement dependent cavity frequency as a function of $\varphi_-$ has the same behavior as shown in Fig.~\ref{fig:freq_I}(c).

Now our Hamiltonian has a similar form to that of the Hamiltonian with symmetric Josephson junctions except for the modified cavity frequency,
\begin{eqnarray}
&&H=\frac{m_r\dot{X}^2}{2}+\frac{m_r\omega_m^2 X^2}{2}+\hbar \omega_c(\varphi_-)a^\dagger a.
\label{Hd}
\end{eqnarray}
The position of the mechanical resonator is quantized by introducing the position operator, which is $X=x_{\text{ZPF}}(b^\dagger+b)$, where $b$ and $b^\dagger$ are creation and annihilation operators and $x_{\text{ZPF}}=\sqrt{\hbar/2m_r\omega_m}$ is the amplitude of zero point fluctuations of the displacement $X$.  Then, the uncoupled Hamiltonian of the mechanical resonator is $\hbar \omega_m b^\dagger b$. 

The interaction terms are obtained by expanding the displacement dependent cavity frequency to the second order in displacement. Then, the interaction Hamiltonian after applying the rotation-wave approximation becomes 
\begin{equation}
H_{int}= \hbar g_{RP}^1  a^\dagger a(b^\dagger +b) +\hbar g^2_{Q} a^\dagger a b^\dagger b,
\end{equation}
where the radiation pressure coupling and cross-Kerr coupling between cavity and mechanical resonator are, respectively,
 \begin{eqnarray}
&&g_{RP}^1=x_{ZPF}\frac{\partial\omega_c}{\partial X}\bigg|_{X=0}=x_{ZPF}\xi\frac{\partial\omega_c}{\partial \varphi_-}\bigg|_{X=0} 
\nonumber
\\
&&\hspace{60pt}=x_{ZPF}\frac{(1-\alpha_I^2)\xi\sin(2\phi_b)\omega_c(0)}{4S(0)^2},\qquad 
\\
&&g^2_{Q}=x_{ZPF}^2\frac{\partial^2\omega_c}{\partial X^2}\bigg|_{X=0}=x_{ZPF}^2\xi^2\frac{\partial^2\omega_c}{\partial \varphi_-^2}\bigg|_{X=0}
\nonumber
\\
&&\hspace{60pt}=2x_{ZPF}\xi g^1_{RP}\cot(2\phi_b)-\frac{3(g^1_{RP})^2}{\omega_c(0)}.\qquad \label{crossKerr}
\end{eqnarray}
Note even when $g_{RP}^1 = 0$, the first term in Eq. (\ref{crossKerr}) stays finite because $\sin(2\phi_b)$ in the radiation pressure coupling is multiplied with the infinite factor $\cot(2\phi_b)$.

\begin{figure}
\centering
    \subfloat[]{\includegraphics[trim = 0 0 130 0, clip, width=\linewidth]{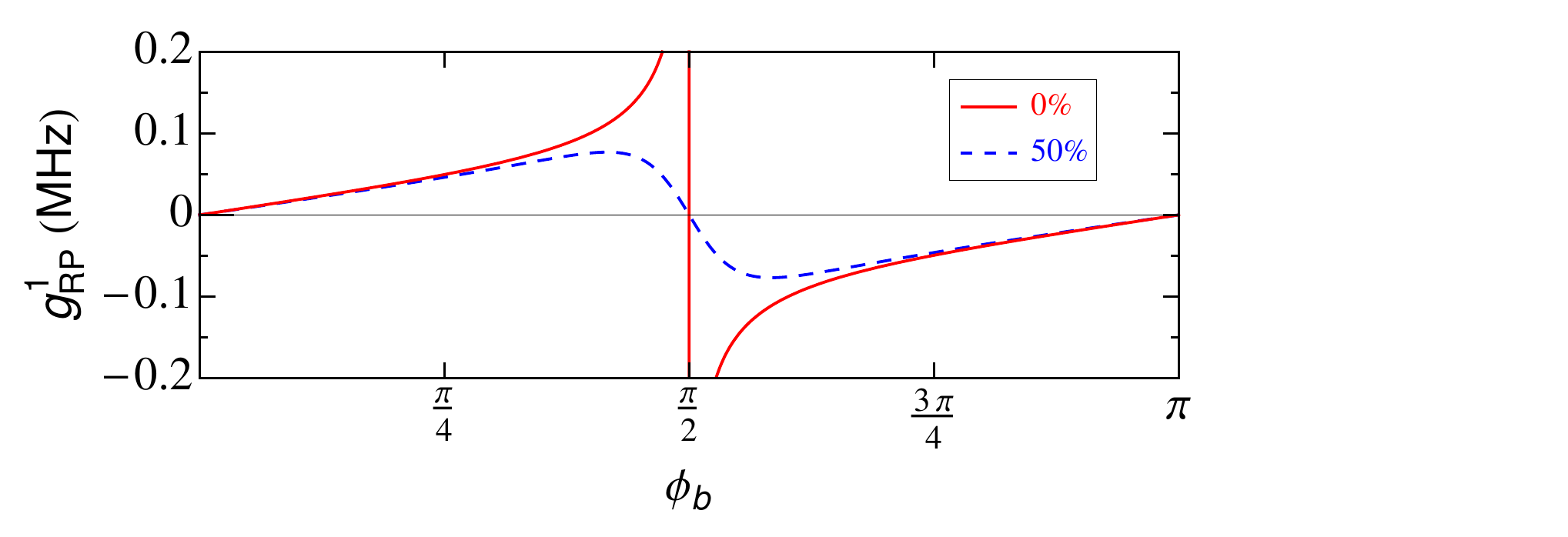}}
    \\
 \subfloat[]{\includegraphics[trim = 0 0 130 0, clip, width=\linewidth]{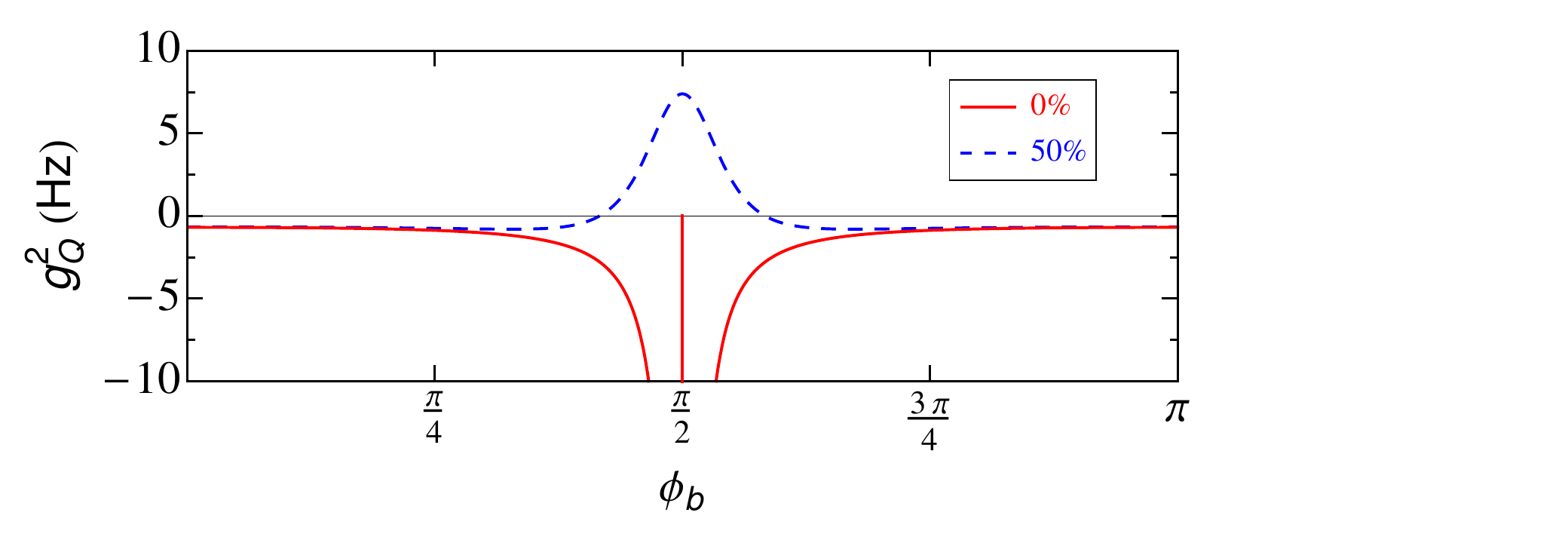}}
  \caption{ Light-matter couplings of symmetric and asymmetric SQUID for magnetic field $B=10$ mT: (a) radiation pressure, (b) cross-Kerr coupling. The maximum of radiation pressure for 50\% asymmetry is  $g_{RP}^1=77$ kHz.The flux bias is shifted by $ \pi B A/\Phi_0=2 \pi 72534$. }
 \label{fig:gdiff}
\end{figure}

\begin{figure}
    \subfloat[]{\includegraphics[ width=0.48\linewidth]{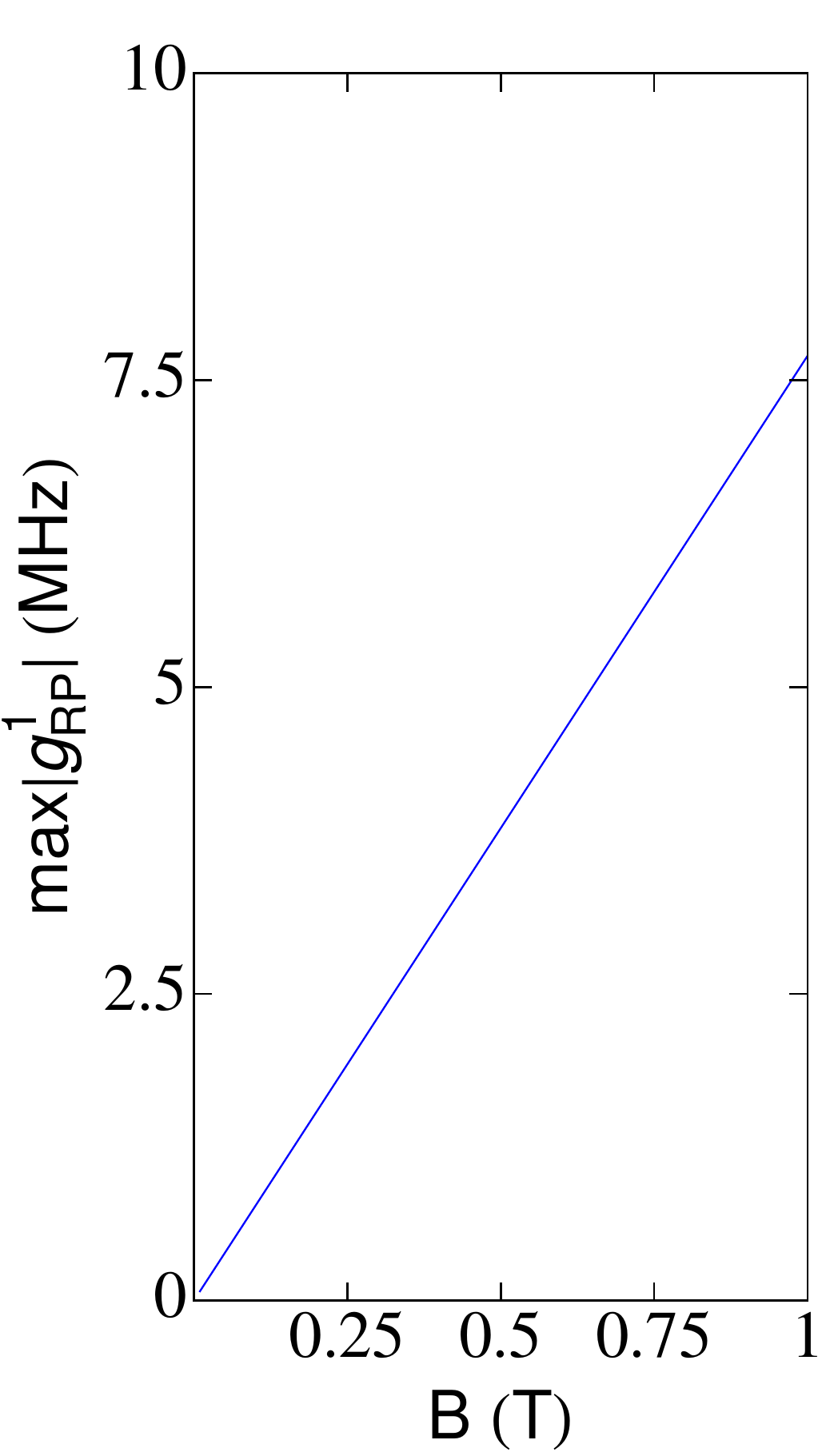}}
    \quad
 \subfloat[]{\includegraphics[ width=0.47\linewidth]{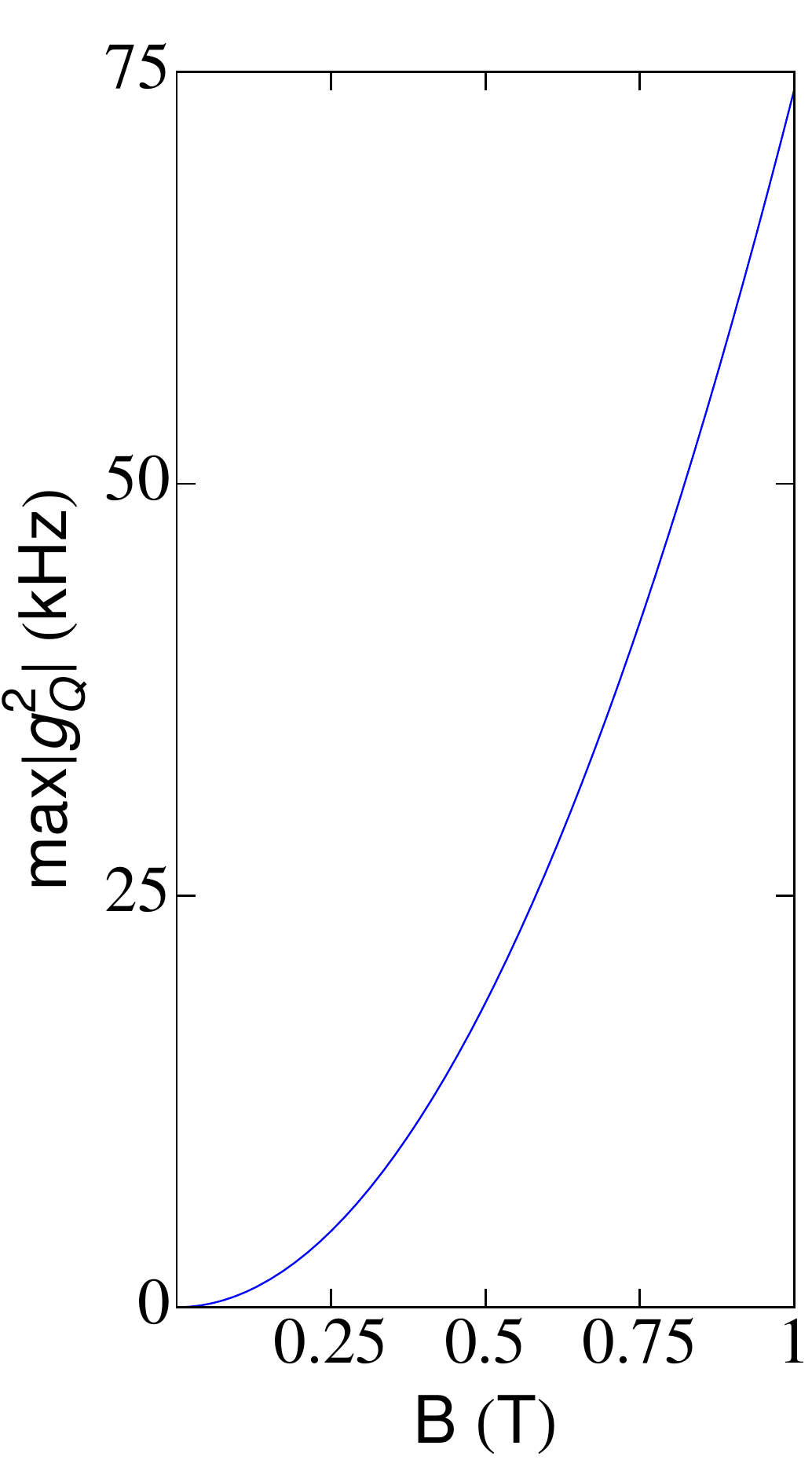}}
  \caption{The maximum of the (a) radiation pressure and (b) cross-Kerr coupling as a function of the magnetic field at 50\% asymmetry. The flux bias is fixed and corresponds to the sweat spot of each coupling.  }
 \label{fig:gdiff_vs_B}
 \end{figure}

In order to visualize the resulting couplings, to the chosen capacitance $C$ and critical current $I_0$ we add following set of parameters: $\omega'_m=10$ MHz, $A=200\hspace{5pt}\mu \text{m}\times 150\hspace{5pt}\mu \text{m}$, $l=150 \hspace{5pt}\mu \text{m}$, and $m_r=200$ pg. The flux bias varies from $\phi_b=2\pi n$ to $\phi_b=2\pi n+\pi$, where $n=72534$ corresponds to chosen value of magnetic field.

In Fig.~\ref{fig:gdiff}(a), we plot the radiation pressure coupling.
For the perfectly symmetric Josephson junctions, the absolute value of the radiation pressure infinitely increases while getting closer to the half-integer flux quantum. It suggests that if in the experiment one can tune bias flux very close to half flux quantum the radiation pressure will be maximum. However, because of the asymmetry of the SQUID the maximum of the radiation pressure coupling shifts to the value of the flux given be  $\tan(\phi_b)=\pm \sqrt{1-\alpha_I^2+\sqrt{1+14\alpha_I^2+\alpha_I^4}}/2\alpha_I$. Behind this value, the radiation pressure monotonically decreases to zero at the half-integer flux quantum. The maximum of the radiation pressure even at 50\% asymmetry  and magnetic field of 10 mT can reach single-photon strong coupling regime, considering a typical cavity decay rate of 80 kHz.

Since the radiation pressure can also be written in terms of the cavity frequency derivative, we can analyze this coupling looking at Fig.~\ref{fig:freq_I}(c) by changing $\phi_b$ to $\varphi_-$ as mentioned above. For the asymmetric case the slope of the frequency increases  and then decreases while changing flux from 0 to $\pi/2$. After crossing $\pi/2$ to $\pi$ it changes sign of the slope, which leads to the negative radiation pressure. Also, for the asymmetric junctions the slope at integer and half integer flux quantum is zero.

The cross-Kerr coupling is shown in Fig.~\ref{fig:gdiff}(b). The coupling $g^2_Q$ is overall much weaker than radiation pressure except for the odd integer flux quantum, where $g^1_{RP}$ is zero and $g^2_Q$ is the leading term.  For the symmetric case the coupling is infinitely strong approaching half-integer flux quantum, which is the same behavior found for the radiation pressure, but in contrast to the latter it does not change the sign while crossing $\pi/2$. Looking at the Fig.~\ref{fig:freq_I}(c) we expect that for the asymmetric  SQUID the cross-Kerr coupling, which is the second derivative of frequency, changes the sign between 0 to $\pi/2$ and then from $\pi/2$ to $\pi$ and this is indeed the result observed here. The maximum of the coupling for the asymmetric case is achieved at the half-integer flux quantum. We can also notice that even at integer flux quantum the value of the cross-Kerr coupling (for the chosen parameters) is $0.66$ Hz. In the experiment with the membrane inside the cavity\cite{Harris_one} the value of the second derivative of the cavity frequency was $\omega_c''(x)/2\pi=108$ kHz nm$^{-2}$. Then in order to improve this value multiple modes of the cavity were coupled to the single mode of the mechanical resonator\cite{Harris_multi} to get $\omega_c''(x)/2\pi=8.7$ MHz nm$^{-2}$, which is still smaller than our calculated value at integer flux quantum, which is  $\omega_c''(x)/2\pi=4$ GHz nm$^{-2}$.

The maximum of the asymmetric couplings increases with magnetic field, which is captured in Fig.~\ref{fig:gdiff_vs_B}. Increasing magnetic field to 1 T is experimentally feasible~\cite{Ben} and increases chances of getting higher couplings. The radiation pressure coupling is linearly dependent on $B$ and the cross-Kerr coupling is quadratically dependent on $B$. At 50\% asymmetry and magnetic field of 1 T the radiation pressure can reach an ulrastrong coupling regime ($g^1_{RP}\sim\omega_m$), which also mean that at lower asymmetry the value on the sweat spot can even be greater.  The cross-Kerr coupling can reach values of 80kHz. It can be stronger for the lower asymmetry, but the window to catch sweat spot becomes more narrow for the lower asymmetry.

\subsection{Resonant frequencies}

In the case of the resonant cavity frequency and mechanical frequency, the minimum of the potential is shifted by the position dependent parameter. However, the displacement is now one of the dynamical variables of the system, separate from the overall phase. If we shift the phase by the displacement dependent parameter then the kinetic energy acquires the shifted phase as well as extra terms in the form of $\dot{\varphi}_+\dot{X}$ with the original phase. Therefore, it is simpler to expand the arctangent in the potential energy  to the first order in $X$, which is sufficient since the amplitude of the mechanical resonator is usually small in such devices. The expanded potential energy depends on both the displacement and the phase $\varphi$, which do not combine in a single variable, 
\begin{eqnarray}
&&E(\varphi,X)=-2E_J S(\varphi_-)\cos\left(\frac{\alpha_I \xi }{S(0)^2}X\right)
\nonumber
\\
&&\qquad\qquad-2E_JS(\varphi_-)\sin\left(\frac{\alpha_I \xi}{S(0)^2} X\right)\varphi
\nonumber
\\
&&\qquad\qquad+E_J S(\varphi_-) \cos\left(\frac{\alpha_I \xi }{S(0)^2}X\right)\varphi^2.
\end{eqnarray}
Similarly to the previous case, the first term shifts the equilibrium position of the mechanical resonator and the mechanical frequency,
\begin{equation}
\omega_m'=\sqrt{\omega_m^2+\frac{2E_J \xi^2\sqrt{2(1+\alpha_I^2+(1-\alpha_I^2)\cos(2\phi_b))}}{m_r}}.
\label{omega_mr}
\end{equation}
Next, we quantize the phase introducing the operators $a^{\dagger}$, $a$. The momentum variable  $p_{\varphi}=m_{\varphi}\dot{\varphi}$ stays the same as in the previous case and the displacement dependent cavity frequency is different from Eq.(\ref{cavity_adiabatic}),
\begin{equation}
\omega_c(X)= \sqrt{\frac{4\pi I_0 S(\varphi_-)  \cos\left(\frac{\alpha_I \xi }{S(0)^2}X\right)}{C\Phi_0}}.
\label{omegaXr}
\end{equation}
Then, the Hamiltonian in terms of the cavity operators has the following form
\begin{eqnarray}
&&H=\frac{m_r\dot{X}^2}{2}+\frac{m_r{\omega'_m}^2 X^2}{2}+\hbar \omega_c(X)a^\dagger a,
\\
&&- \frac{2\hbar I_0 S(\varphi_-)}{\sqrt{2\hbar C\omega_c(X)}}\sin\left(\frac{\alpha_I \xi}{S(0)^2} X\right)(a^\dagger+a).
\label{Hr}
\end{eqnarray}

We expand the full Hamiltonian to the second order in the displacement and use the creation and annihilation operators $b^{\dagger}$, $b$ of the mechanical resonator. It leads to the uncoupled cavity Hamiltonian $\hbar \omega_c(0)a^\dagger a$  and the uncoupled mechanical resonator Hamiltonian $\hbar \omega'_m b^\dagger b$. Applying the rotating-wave approximation results in the interaction Hamiltonian of the following form
\begin{equation}
H_{int}=\hbar g^{2r}_{Q} a^\dagger a b^\dagger b -\hbar g^{1r}_{BS}(a^\dagger b+ b^\dagger a),
\end{equation}
where the cross-Kerr and the single-photon beam splitter couplings, respectively, are
\begin{eqnarray}
&&g^{2r}_{Q}=x_{ZPF}^2\frac{\partial^2\omega_c^r}{\partial X^2}\bigg|_{X=0}=g^{2}_{Q}-\frac{\alpha_I^2\xi^2}{2S(0)^4}\omega_c(0),
\label{g2r}
\\
&&g^{1r}_{BS}=x_{ZPF}\frac{\alpha_I\xi \sqrt{\omega_c(0) E_J}}{S(0)\sqrt{\hbar S(0)}}.
\end{eqnarray}

To plot these couplings, we use set of parameters for the nanoSQUID with CNT junctions\cite{CNT-SQUID}: $I_0=15 $ nA, $C=90$pF, $A=800\hspace{5pt} \text{nm}\times 800\hspace{5pt} \text{nm}$, $l=200 \hspace{5pt} \text{nm}$, and $m_r=5$ ag. The cavity frequency for this values is 1 GHz. The mechanical frequency is taken to be  $\omega'_m=1$ GHz, which is possible to reach with a suspended CNT. The flux bias varies from $\phi_b=2\pi n$ to $\phi_b=2\pi n+\pi$, where $n=18$.

\begin{figure}
\centering
 \subfloat[]{\includegraphics[trim = 0 0 130 0, clip, width=\linewidth]{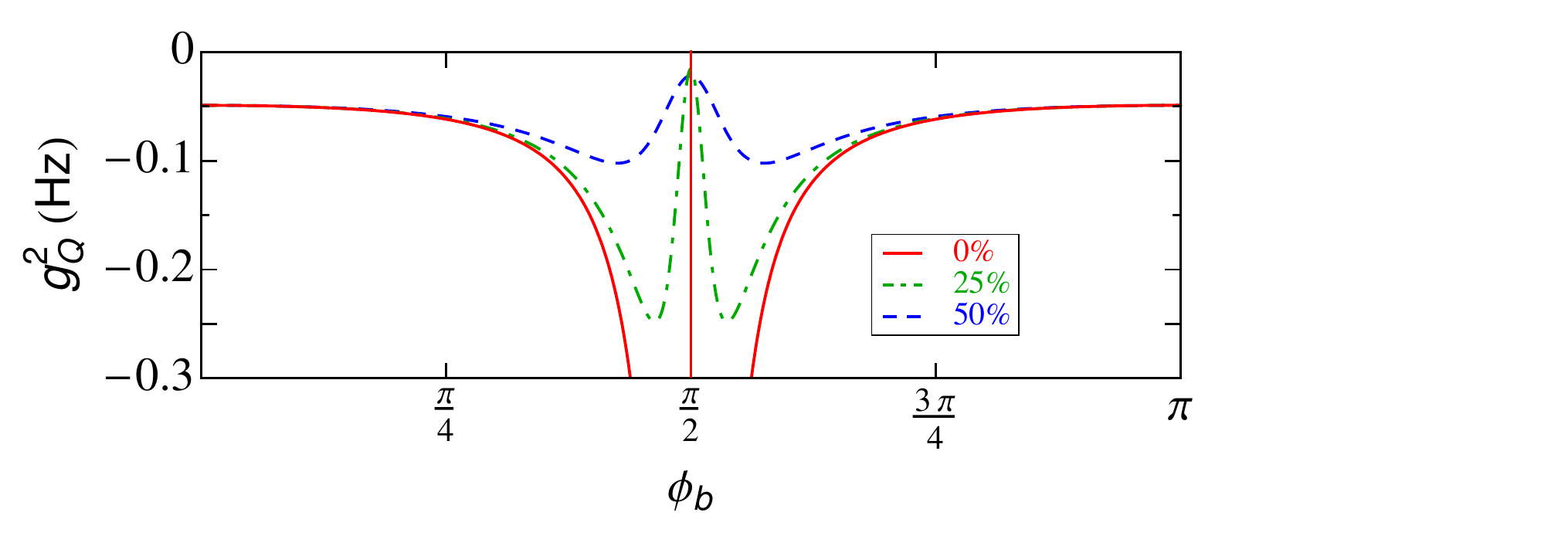}}
\\
    \subfloat[]{\includegraphics[trim = 0 0 130 0, clip, width=\linewidth]{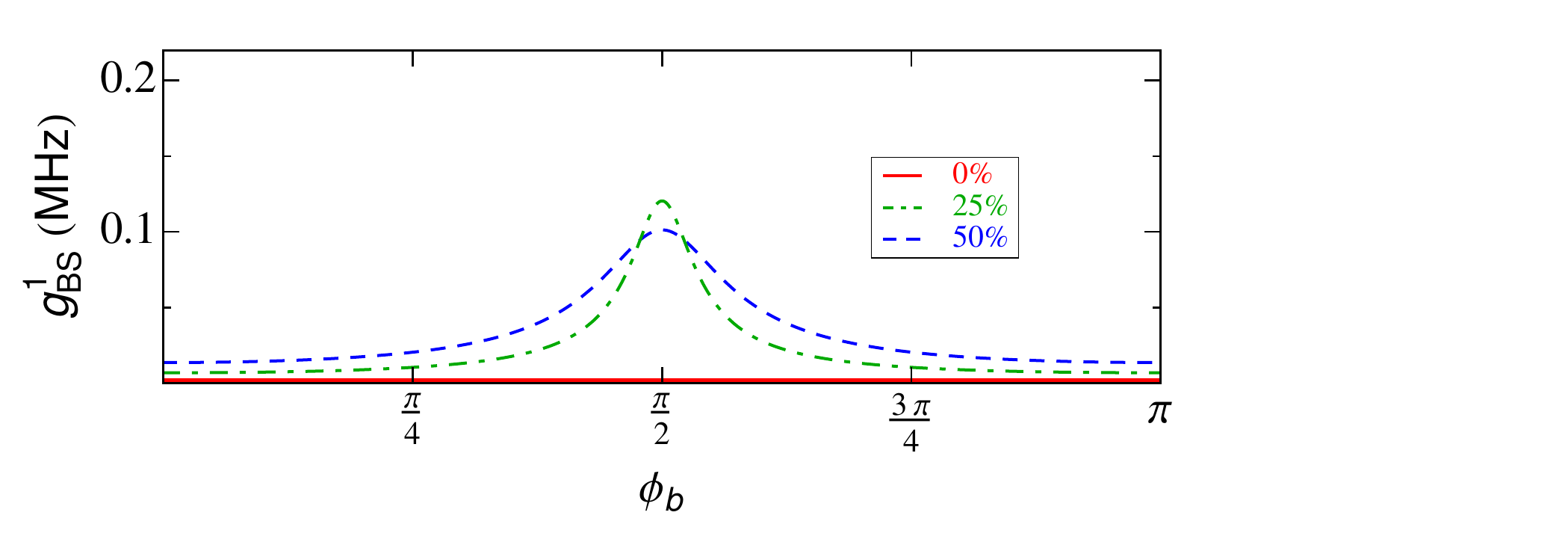}}
  \caption{Interaction couplings for symmetric and asymmetric SQUID in the case of $\omega_c(0)\sim\omega'_m$ and magnetic field $B=10$ mT: (a) cross-Kerr coupling, (b) single-photon beam splitter coupling. The flux bias is shifted by $ 12 \pi B A/\Phi_0=2 \pi 18$. }
 \label{fig:gdiffr}
\end{figure}

\begin{figure}
    \subfloat[]{\includegraphics[ width=0.48\linewidth]{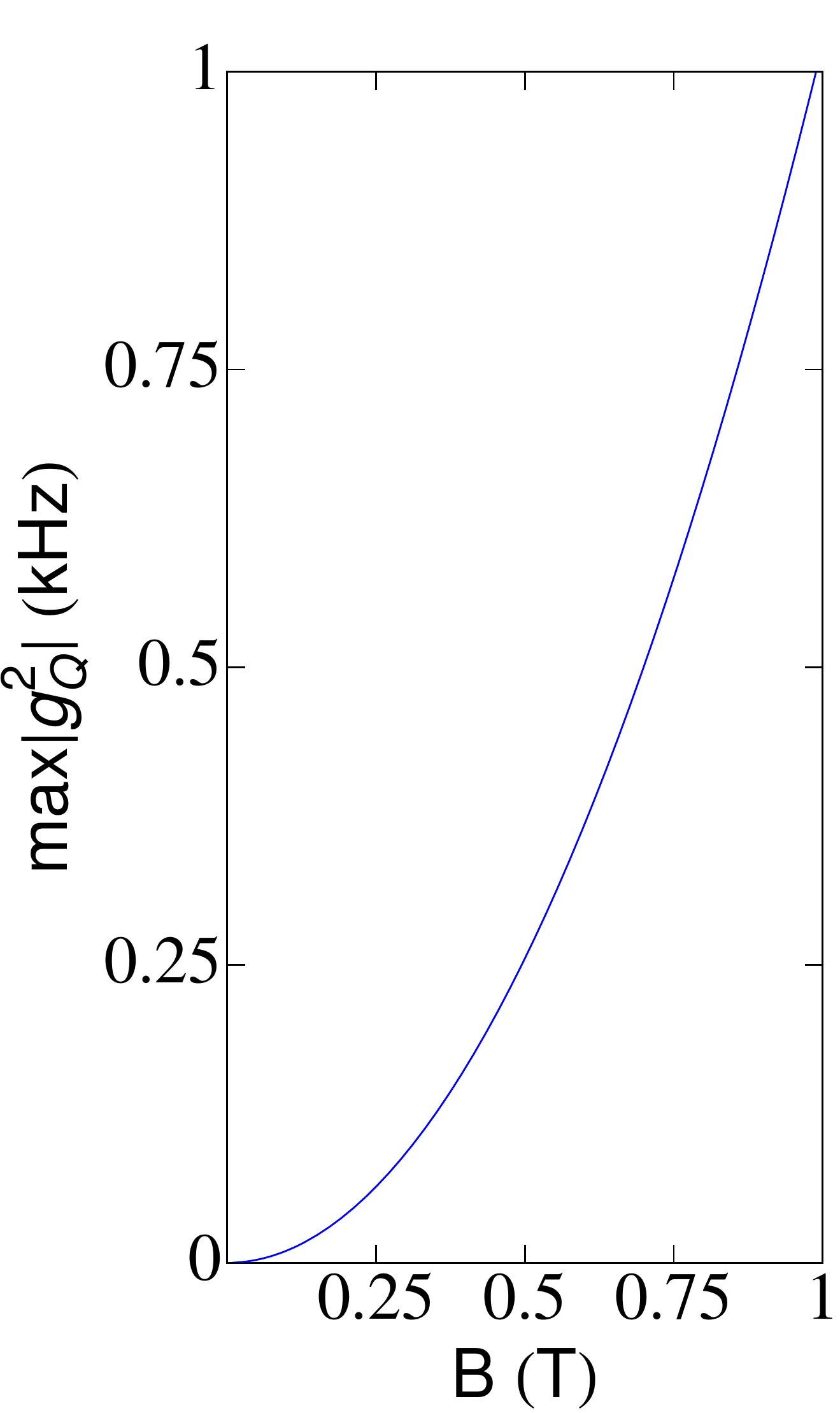}}
    \quad
 \subfloat[]{\includegraphics[ width=0.47\linewidth]{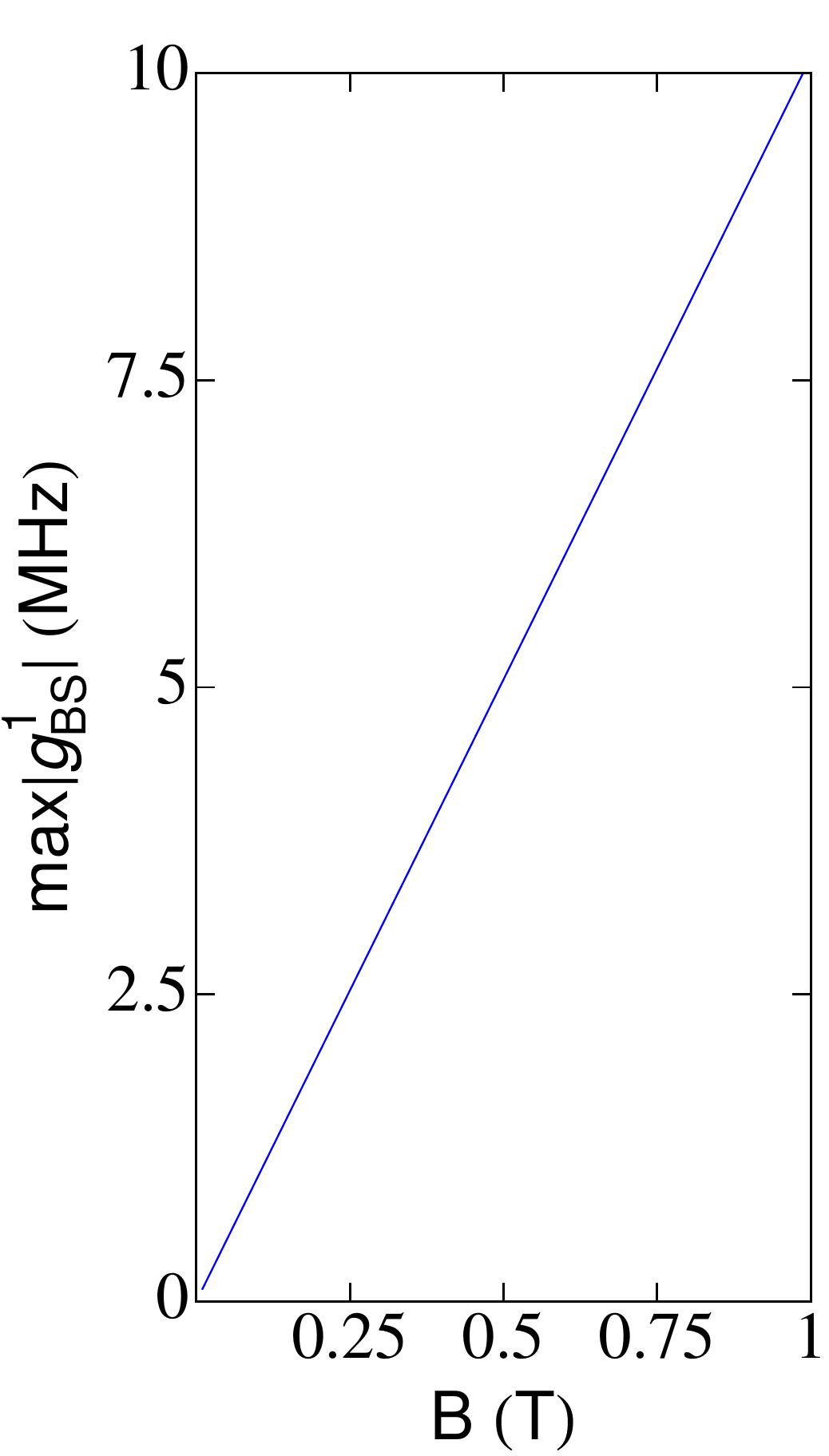}}
  \caption{The maximum of (a) the cross-Kerr coupling and (b) the single-photon beam splitter coupling at 50\% asymmetry, as a function of the magnetic field corresponding to Fig.~\ref{fig:gdiffr}. The flux bias is fixed to the sweet spot  of each coupling.  }
 \label{fig:gdiffr_vs_B}
 \end{figure}

In Fig.~\ref{fig:gdiffr}(a), we plot the cross-Kerr coupling. The coupling $g^2_Q$ is overall weaker than the one we found in Fig.~\ref{fig:gdiff}(b). For the symmetric case, the behaviour is the same as we have seen in the dispersive regime. However, for finite asymmetry the behavior is qualitatively different since the coupling is negative and does not change the sign from negative to positive. Also, there is a peak of the coupling close to $\pi/2$ and at exactly half flux quantum there is a minimum. This happens because of the second term in Eq.\eqref{g2r}, which arise due to the modified cavity frequency. For lower asymmetry the minimum is getting closer to zero while the maximum is increasing.

Fig.~\ref{fig:gdiffr}(b) shows the single-photon beam splitter coupling. For $\omega_c(0)\gg\omega'_m$ this coupling is negligible, because the corresponding term in the Hamiltonian is a quickly oscillating function of time. It only exists at a finite asymmetry. The single-photon beam splitter coupling is slowly increasing while flux rises from 0 to $\pi/2$ and reaches its maximum at the half flux quantum and then passing this point decreases again. For lower asymmetry the peak is higher, but the window to reach higher value is narrower, since the higher asymmetry corresponds to a higher value of the coupling except for close to half-integer flux quantum.

In Fig.~\ref{fig:gdiffr_vs_B} we show the maximum of both couplings at the 50\% asymmetry while increasing magnetic field to 1 T.  The cross-Kerr coupling reaches 1 kHz. The single-photon beam splitter coupling has the value up to 10 MHz. In comparison with the mechanical frequency it is not in the ultrastrong coupling regime, but depending on the cavity decay rate it can be in the strong-coupling regime. While often radiation pressure is linearized to produce the beam splitter term to solve specific systems and phenomenon, $g^{2r}_{BS}$ has intrinsically a beam splitter character even at the single-photon level. At very small asymmetry and strong magnetic fields the cross-Kerr coupling can be larger than beam-splitter interaction.

\section{Discussion}
\label{sec:dis}
\begin{figure}
 \includegraphics[ width=0.95\linewidth]{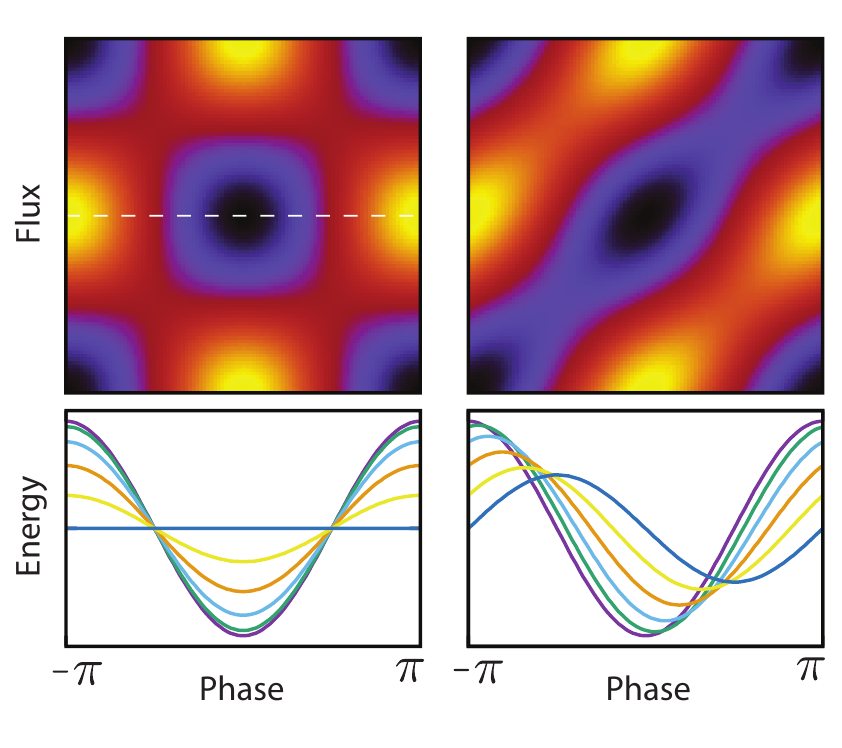}
  \caption{ Potential energy as a function of the phase and flux at 0\% (on the left) and 50\% asymmetry(on the right). The minimum is shown in black and maximum in yellow. The corresponding cross section of energy map for the different flux is displayed on the bottom.   }
 \label{fig:pot_energy}
\end{figure}

To gain the intuition about the couplings and to better understand the asymmetric system we plot the potential energy as a function of flux ($\varphi_-$) and phase ($\varphi_+$) in Fig.~\ref{fig:pot_energy}. For the symmetric junctions the potential is symmetric along dashed line. At the bottom of the potential radiation pressure is zero and cross-Kerr coupling is finite. For the asymmetric case the potential has an elliptical form and is asymmetric. One can further study the figures on the bottom corresponding to the cross-section of the energy map for different flux. The cross-sections are chosen by moving the horisontal dashed line up. In the symmetric case the minimum of the potential energy stays at the same position. In contrast, for the asymmetric junctions the position of the minimum  shifts, which is also described by Eq.\eqref{Energy}. In the case of dispersive frequencies, this shift it constant and the minimum is redefined at each bias flux, which represents the same physics as for the symmetric case. However, elliptical form  alters the  radiation pressure and cross-Kerr couplings and changes its behavior as previously appeared in Fig.~\ref{fig:gdiff} around half-integer flux quantum where there is a merge between elliptical forms. 

For the case of resonant frequencies and asymmetric SQUID, the minimum of the potential energy is shifted by the displacement dependent flux. The oscillations of the mechanical resonator correspond to the  motion from one curve to another one. The force that triggers the motion between the minima is just like the Lorentz force, which explains the appearance of the single-photon beam splitter and the extra term picked up by cross-Kerr coupling as compared to the dispersive regime. 

We now discuss the mechanical frequency shifts due to the Josephson term in Eqs.\eqref{omega_m} and \eqref{omega_mr}. In the parameter regime we have chosen this shift can be disregarded. However, for higher Josephson energy, which could depend on the magnetic field itself, or larger parameters of the mechanical resonator increasing magnetic field to 1 T can create a large shift, which should be taken into account. The mechanical zero-point fluctuation in such situation becomes smaller, subsequently the values of the couplings decrease, but we do not expect the change in the overall behavior of the optomechanical couplings.

Generally, the SQUID is an intrinsically nonlinear cavity and close to the half integer flux quantum an extra nonlinear Kerr-type term $\Lambda a^\dagger aaa^\dagger$ appears in the Hamiltonian of Eqs. \eqref{Hd} and \eqref{Hr}, where the Kerr nonlinearity is $\Lambda=\hbar \pi^2/(4 C \Phi_0^2)$. This term results from the expansion of the potential energy to the forth order in the overall phase $\varphi$. Thus, a cavity can be considered linear as long as $\Lambda$ is less than the cavity linewidth and $\omega_c(0)\gg \Lambda$, which gives a finite condition for the flux bias close to the half integer flux quantum. Close to the half flux quantum the Kerr-type term and the cross-Kerr term are always present in this system. From the fourth order expansion of the potential energy there are also other nonlinear interaction terms such as $a^\dagger a^\dagger aa b^\dagger b$ in the dispersive case or  $a^\dagger a^\dagger a b$ in the resonant case, which are always small.

\section{Conclusions}
\label{sec:concl}

We provided a quantum analysis of the SQUID with asymmetric Josephson junctions and embedded mechanical resonator for two cases of the dispersive and resonant cavity and mechanical frequencies. Our findings are significant for the experimental setup where asymmetry cannot be avoided.  We found that the radiation pressure for the resonant frequencies has a sweet spot, which  is located at an asymmetry dependent flux point. Shifting this point towards the half-integer flux quantum results in a weaker coupling. Even at 50\% asymmetry and weak magnetic field the radiation pressure coupling can be in the strong coupling regime. For high magnetic fields, the ultrastrong coupling regime of the radiation pressure can be achieved. The cross-Kerr coupling is finite at the odd integer flux quantum, in contrast to the radiation pressure coupling. For the symmetric case, it is always negative and infinitely strong very close to the half-integer flux quantum. For the dispersive asymmetric case, the cross-Kerr coupling has maximum at half-integer flux quantum and changes sign from negative to positive while reaching maximum. For the resonant asymmetric case, the minimum sits at the half-integer flux quantum and maximum is at the flux dependent point close to $\pi/2$. For the resonant case, the radiation pressure is too weak since it oscillates at higher frequency and instead single-photon beam splitter interaction is the main term in the Hamiltonian. It is always finite and has its maximum at the half-integer flux quantum. 

We explained the origin of different couplings using the potential energy map as well as compared the maps for the symmetric and asymmetric cases.  The biggest challenge to experimentally work with single-photon beam splitter coupling is the condition on the mechanical frequency, which should be comparable to the cavity frequency. Experiments involved such setup previously\cite{vanderZant,Etaki,Etaki1} had  mechanical frequency smaller than cavity frequency. However, using carbon nanotubes  as  a mechanical resonator coupled to the Josephson circuit can potentially solve the high mechanical frequency issue. 

\section*{Acknowledgments}

This work was supported by the Netherlands Foundation for Fundamental Research on Matter (FOM).

\end{document}